\documentclass[fleqn,usenatbib]{mnras}

\usepackage{newtxtext,newtxmath}

\usepackage[T1]{fontenc}

\DeclareRobustCommand{\VAN}[3]{#2}
\let\VANthebibliography\thebibliography
\def\thebibliography{\DeclareRobustCommand{\VAN}[3]{##3}\VANthebibliography}

\newcommand{\Msun}{\,{\rm M_\odot}}

\def\SFRUV{\mathrm{SFR}_{\mathrm{UV}}}
\def\LUV{\mathrm{L}_{\mathrm{UV}}}
\newcommand{\pushright}[1]{\ifmeasuring@#1\else\omit\hfill$\displaystyle#1$\fi\ignorespaces}
\def\be{\begin{equation}}
\def\ee{\end{equation}}
\def\ba{\begin{align*}}
\def\ea{\end{align*}}

\usepackage{graphicx}	
\usepackage{amsmath}	
\usepackage{amssymb}	
\usepackage{pgfplots}
\usepackage{subfig}
\usepackage[justification=centering]{caption}
\usepackage{pbox}

\hypersetup{colorlinks=true}

\usepackage{tipa}
\pgfplotsset{width=6.6cm,compat=1.7}  
\title[Lensing Hypothesis for $z \sim 13$ Galaxies]{The Two $z\sim 13$ Galaxy Candidates HD1 and HD2 Are Likely Not Lensed}

\author[R. Z. Lee et al.]{
Rui Zhe Lee$^{1}$\thanks{rlee@college.harvard.edu},
Fabio Pacucci$^{1,2}$\thanks{fabio.pacucci@cfa.harvard.edu}, Priyamvada Natarajan$^{2,3,4}$ \& Abraham Loeb$^{1,2}$
\\
$^{1}$Center for Astrophysics $\vert$ Harvard \& Smithsonian, Cambridge, MA 02138, USA\\
$^{2}$Black Hole Initiative, Harvard University,
Cambridge, MA 02138, USA\\
$^{3}$Department of Astronomy, Yale University, 52 Hillhouse Avenue, New Haven, CT 06511, USA\\
$^{4}$Department of Physics, Yale University, P.O. Box 208121, New Haven, CT 06520, USA
}

\date{\today}

\pubyear{2022}

\begin{document}
\label{firstpage}
\pagerange{\pageref{firstpage}--\pageref{lastpage}}
\maketitle

\begin{abstract}
The discovery of two UV-bright galaxy candidates at $z\sim 13$, HD1 and HD2, laid the foundation for a new race to study the early Universe. Previous investigations suggested that they are either powered by a supermassive black hole or by an extreme, transient burst of star formation. Given their uncertain nature, we investigate whether these sources could be lensed by a hitherto undetected, faint foreground galaxy. We find that at the current limiting magnitude with which HD1 and HD2 were imaged, there is only a $7.39\%$ probability they are strongly lensed by spherical deflectors and that the hypothetical lensing galaxy was too faint to be detected. Meanwhile, with the limiting magnitudes of the Hubble Space Telescope (HST) and James Webb Space Telescope (JWST), the theoretical probability would drop precipitously to $0.058\%$ and $0.0012\%$, respectively. We further find it unlikely that the luminosities of both sources can be accounted for by lensing that produces a single, resolved image with sufficiently high magnification. Alternatively, in the unlikely event that their brightness results from lensing by an elliptical isothermal galaxy, there is a $30.9 \%$ probability that the lensing galaxy is too faint to be observable at the current limiting magnitude. Future HST (JWST) imaging will drop this probability to $0.245 \%$ ($0.0025 \%$). In summary, while deep imaging with HST and JWST is required to discard the lensing hypothesis entirely, it is unlikely that the exceptional luminosity of the two $z \sim 13$ sources can be accounted for by gravitational lensing.
\end{abstract}




\section{Introduction}

A pair of H-band drop-out Lyman Break Galaxy (LBG) candidates, HD1 and HD2, have been recently discovered, placing them at a photometric redshift of $z \sim 13$ \citep{Harikane_2022_LBG}. This prior detection is consistent with the large number of recent observations from the James Webb Space Telescope (JWST) of galaxy candidates at $z > 12$, and with the suggestion that massive galaxy formation could have had a very early onset, with galaxies as massive as $\sim 5\times10^9 \Msun$ in stars possibly in place at $z\sim 18$ \citep{Labbe_2022}.

\cite{Pacucci_2022} showed that at the inferred number density of these $z \sim 13$ candidates, HD1 and HD2, their UV magnitudes $(\mathrm{M}_{\mathrm{UV}} \sim - 23.5)$ are unusually bright. When compared to theoretical predictions from star-formation models extrapolated from $z < 10$, they appear to be over-luminous outliers $\sim 2-3$ magnitudes brighter than expectations (see, e.g., \citealt{Finkelstein_2015, Oesch_2018, Bowler_2020, Bouwens_2021}). 

Some explanations provided to account for this discrepancy include: (i) that the sources are extreme star-forming galaxies with star formation rates (SFRs) several times higher than expected for typical galaxies at a given halo mass \citep{Harikane_2022_Goldrush}, possibly, due to an ongoing transient starburst; (ii) that the star formation efficiency generally increases with halo mass, but this effect is compensated by increasing dust attenuation at $z \gtrsim 10$, which prevents evolution at the bright end of the UV luminosity function (LF) \citep{Harikane_2022_Goldrush}; (iii) that the initial mass function (IMF) of their stellar population is more top-heavy compared to the local IMFs \citep{Harikane_2022_LBG, Pacucci_2022}; and (iv) a significant fraction of the detected luminosity is generated by accretion onto $\sim 10^8 \Msun$ supermassive black holes hosted in HD1 and HD2 \citep{Pacucci_2022}. 

Implicit in these studies, however, is the assumption that the observed UV magnitudes and associated luminosities are intrinsic, i.e., the galaxies are not lensed by foreground mass distributions along the line of sight. Depending on the level of alignment, gravitational lensing by intervening foreground deflectors can result in one or more magnified and distorted images of the distant source in the sky, accompanied by an increase in the brightness of the source, as surface brightness is conserved (see \citealt{Schneider_1992}, \citealt{Narayan_1996}, \citealt{KneibPN2011} for lensing basics). For example, the only quasar thus far detected in the epoch of reionization at $z > 6$, J0439+1634, was originally interpreted as extremely bright, before the discovery that it was lensed and magnified by a factor $\mu \approx 50$ \citep{Fan_2019, Pacucci_2019}. Our ultimate goal is to infer the physical properties of these extremely high-redshift sources, HD1 and HD2, from their observed UV magnitudes. Even though the sources appear to be spatially isolated with no evident sign of lensing \citep{Harikane_2022_LBG}, it is still imperative to consider the possibility that they are lensed by foreground deflectors.

The lack of any gravitational lensing effect by eye inspection could be due to low-resolution imaging available at this time \citep{Harikane_2022_LBG}. Additional techniques could be used as well to determine the lensing likelihood. For example, sub-mm selection of galaxies has proven to be a very efficient method to sift out lensed galaxies. In fact, due to the extremely steep number counts predicted for $z > 1$ sub-mm galaxies, only strongly lensed sources can be detected at $\sim 500 \, \mathrm{\mu m}$, with sub-mm fluxes $\gtrsim 100$ mJy (see, e.g., \citealt{Negrello_2007, Wardlow_2013, Negrello_2017}). Additional, multi-wavelength observations of HD1 and HD2 will be fundamental to assess the possibility that they are gravitationally lensed.

\cite{Harikane_2022_Goldrush} already suggested that the bright end of the luminosity function of galaxies at $z=2-7$ could be explained by gravitational lensing. In fact, they show that a lensed version of the Schechter function provides a better fit to data than the unlensed one. 
Unfortunately, in their analysis, they also point out that a visual inspection of their sources does not show significant evidence for magnifications and/or elongated morphologies.

\cite{Harikane_2022_LBG} was the first study to report the discovery, via photometric fitting, of $z\sim 13$ galaxies. Since then, a large wealth of new, extremely high-$z$ photometric candidates based on JWST data has been published
\citep{Adams_2022, Castellano_2022, Donnan_2022,Finkelstein_2022, Yan_2022, Naidu_2022, Cullen_2022}. Once the redshift of these novel JWST candidates is ultimately established, it will be crucial to determine if some of them are also affected by gravitational lensing.

In this study, we consider various lensing scenarios, and place constraints on the properties of such putative deflectors. In particular, we also consider the effects of naked cusps, which are a rare type of lensing configuration that possibly lead to highly magnified images, when simulating the lensing effects of elliptical deflectors in Section 4. With the methods employed in this study, however, we are not able to consider the possibility of dark lenses, which might serve to magnify HD1 and HD2 sufficiently while remaining hidden from direct observations. We use the Planck Collaboration 2018 cosmology \citep{Planck_2018} for all calculations. For brevity, we only compute in detail the consequences of the lensing hypothesis for HD1 at an inferred redshift of $z \approx 13.27$, as the results of our study hold equally well for HD2, at a comparable redshift of $z \approx 12.3$ \citep{Harikane_2022_LBG}.
As a reminder, the observed UV luminosity and SFR for HD1 are $\LUV = 10^{11.7} \, \mathrm{L_{\odot}}$ and $\SFRUV = 78.9 \Msun \, \mathrm{yr}^{-1}$, while for HD2 they are $\LUV = 10^{11.9} \, \mathrm{L_{\odot}}$ and $\SFRUV = 125.2 \Msun \, \mathrm{yr}^{-1}$, respectively.

\section{Strong Gravitational Lensing with a Single Magnified Image}

In gravitational lensing, the properties of the observed images are determined primarily by the relative alignment of source and lens, hence by the Einstein radius, $\theta_E$, of the configuration. In general, $\theta_E$ depends on the intrinsic properties of the lensing galaxy, primarily its mass distribution, and ellipticity. The contribution of additional masses in the proximity of the primary lens can induce a tidal shear, that acts as an additional perturbation \citep{Schneider_1992}.
In addition, $\theta_E$ depends on cosmological parameters via the geometric factors of the lensing configuration through $D_S, D_L \ \mathrm{and} \ D_{LS}$, which are the angular diameter distance to the source, the lens, and between the source and the lens, respectively. 
In the strong gravitational lensing regime, wherein the magnification exceeds unity, lensed background galaxies appear multiply imaged and strongly distorted with some images magnified and others de-magnified. In solving the lens equation, critical curves can be derived in the lens plane along which images experience maximal magnification. These critical curves can be mapped onto caustics in the source plane which divide the source plane into regions of different image multiplicity. In the image plane of a lensing system, the magnification of an image is formally infinite at the critical curves. In general, the number of images produced by the lens mapping changes every time the source position crosses a caustic \citep{Schneider_1992, Narayan_1996}.

We first consider the possibility that the detected $z \sim 13$ sources HD1 and HD2 lie outside the caustics of their foreground lensing galaxies, are singly-imaged, and magnified. We estimate the magnification factor $\mu$ that their single images\footnote{Strong lensing is customarily defined to be the regime in which both the convergence $\kappa > 1$ and multiple images are produced. Here, since we are dealing with the regime where no multiple images are produced while $\kappa > 1$, we use the term strong lensing with this clarification.} will need to be magnified by in order to reconcile their inferred intrinsic luminosities with that of models extrapolated from lower redshifts. First, we consider the copiously star-forming galaxy hypothesis, \cite{Pacucci_2022} and use the relation $\SFRUV = K \, \LUV$ where $K \approx 8.23 \times 10^{-29} \ \Msun \  \mathrm{yr}^{-1} \ \mathrm{erg}^{-1} \ \mathrm{s \ Hz}$ to compare the intrinsic SFR of the sources to that derived from scaling relations with star formation rates  from lower redshifts. The factor $K$ is derived in \cite{Pacucci_2022} and expressed the amount of UV luminosity produced by a unit of star formation rate. Since the magnification $\mu$ is the ratio of the observed flux over the intrinsic flux, we can relate the intrinsic SFR to $\mu$ using the relation above:
\begin{equation} \label{eq1}
    \SFRUV = K \frac{\mathrm{L}_{\rm obs}}{\mu} \, .
\end{equation}
Here, $\mathrm{L}_{\rm obs} = 10^{11.7} \, \mathrm{L_{\odot}}$ for HD1 and $\mathrm{L}_{\rm obs} = 10^{11.9} \, \mathrm{L_{\odot}}$ for HD2, both at the UV restframe wavelength of $\lambda = 1500$ \AA \ \citep{Pacucci_2022}.

We can obtain $\mu$ by turning to the specifics of the lensing configuration. For simplicity, we assume that the lensing galaxy can be described by an SIS (Singular Isothermal Sphere) lens profile. For this profile, lensing with a single magnified image and no multiple images is achieved when the source lies outside the caustics \citep{Schneider_1992}. Equivalently stated, the deflection angle subtended by the source to the center of the symmetric lens from the perspective of the observer $(\beta)$ is greater than the Einstein radius, i.e., for $\beta > \theta_E$. The Einstein radius for such an SIS lens is given by: 
\begin{equation} \label{eq2}
    \theta_E = 4 \pi \left( \frac{\sigma}{c}\right)^2\frac{D_{LS}}{D_S} \, ,
\end{equation}
where $\sigma$ is the line-of-sight central velocity dispersion of the galaxy, which is directly related to the mass of the galaxy. Using $D_{LS}=(x_S-x_L)/(1+z_S)$, we can rewrite Eq. \ref{eq2} as:
\begin{equation} \label{eq3}
    \theta_E = 4 \pi \left( \frac{\sigma}{c}\right)^2\frac{x_S - x_L}{x_S} \, ,
\end{equation}
where $x_L$ and $x_S$ are the corresponding comoving distances to the lens and source respectively \citep{Hogg_1999}. Using $\theta_E$ from Eq. \ref{eq3}, we calculate the magnification $\mu$ of the single image from the deflection angle $\beta$ as
\begin{equation} \label{mu}
    \mu = 1+ \theta_E/\beta \, .
\end{equation}
We note that this equation only holds in the regime where $\beta > \theta_E$ on the image plane. Using the above expressions, we calculate the variation in the derived SFR against the degree to which the source is magnified, which in turn depends on $\beta$.

Since the lensing properties of the foreground galaxy modeled as an SIS can be well described by the velocity dispersion $\sigma$ and its redshift $z_L$, we calculate the variation in SFR against $\beta$ across a range of values for $\log \sigma$ and $z_L$. We note that typical velocity dispersions for galaxies in units of $\mathrm{km} \ \mathrm{s}^{-1}$ falls in the range $1.6 \leq \log \sigma \leq 2.6$ \citep{Sohn_2017, Sohn_2020}, and the lensing optical depth of any arbitrary source, including all lensing configurations considered in this study, drops off to negligible levels after $z_L \sim 3$ \citep{Wyithe_2011, Mason_2015, Pacucci_2019}. 

\begin{figure}
\includegraphics[width=\columnwidth]{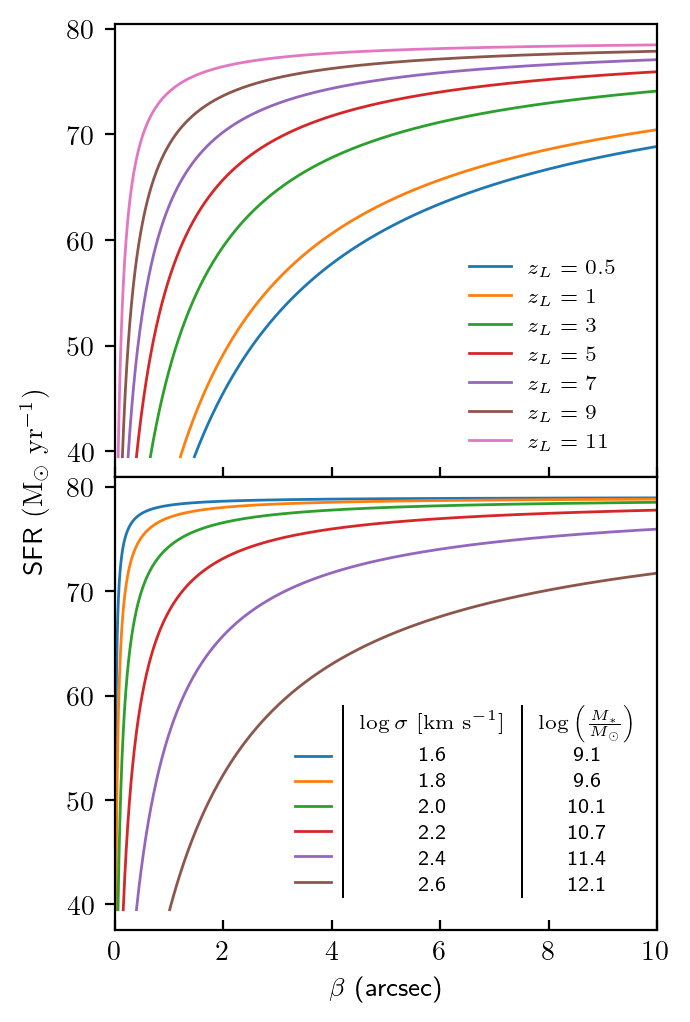}
\centering
\caption{\textbf{Top:} SFR against $\beta$ for fixed $\log \sigma = 2.4$ and $M_\star \sim 10^{11.4} \Msun$, across varying values of $z_L$. \textbf{Bottom:} SFR against $\beta$ for fixed $z=5$, across varying values of $\log \sigma$, or equivalently, $M_\star$.}
\label{fig:1}
\end{figure}

Furthermore, to obtain insights on what the stellar mass of a galaxy at a given $\sigma$ is likely to be, we use the empirically derived $M_\star-\sigma$ relation \citep{Zahid_2016}. The variation in SFR against $\beta$ for various values of $z_L$ and fixed $\log \sigma = 2.4$, corresponding to a galaxy stellar mass of $M_\star \sim 10^{11.4} \Msun$, is displayed in the top panel of Fig. \ref{fig:1}. The variation in SFR against $\beta$ for various $\log \sigma$ and fixed lens redshift $z_L$ = 5 is, instead, shown in the bottom panel of Fig. \ref{fig:1}, along with the galaxy mass corresponding to each $\log \sigma$ shown alongside. Note that in both plots, only values of $\beta > \theta_E$, the condition that needs to be satisfied to produce a single, magnified lensed image are displayed. 

For this lensing configuration, the minimum intrinsic SFR for HD1 derived for all possible magnified, yet singly-imaged, lensing scenarios across the parameter space is $\gtrsim 40 \Msun \, \mathrm{yr}^{-1}$. Compared to the estimated intrinsic SFR of $5.3 - 16.5 \Msun \, \mathrm{yr}^{-1}$ from the expected number density of galaxies at $z \sim 13$ \citep{Harikane_2022_LBG, Pacucci_2022}, we see that the possibility of HD1 being lensed by an SIS deflector producing a single image cannot account for its unusually high observed luminosity.
The observed UV luminosity and SFR for HD1 are $\LUV = 10^{11.7} \, \mathrm{L_{\odot}}$ and $\SFRUV = 78.9 \Msun \, \mathrm{yr}^{-1}$, respectively.
We further note that by following the calculations above, we find that the same conclusion holds for HD2, which has an observed UV luminosity and SFR of $\LUV = 10^{11.9} \, \mathrm{L_{\odot}}$ and $\SFRUV = 125.2 \Msun \, \mathrm{yr}^{-1}$.

We stress that the above conclusion only holds for the assumption of the lensing galaxy modeled as a spherically symmetric SIS  model. Therefore, we need to consider an elliptic mass model for the deflector next. Indeed, ellipticity in the total mass distribution of the primary lensing galaxy, as well as gravitational tidal shear from nearby objects along the line of sight will necessarily lead to non-spherical lensing cross sections, potentially also producing single images with magnifications $\mu \gg 2$ \citep{Keeton_2005}. This occurs when the source is outside but still relatively close to the outer caustic on the source plane. However, this effect has been demonstrated to not be very significant, since for typical values of ellipticity $(\epsilon \sim 0.3)$ and shear $(\gamma \sim 0.1)$, the maximum magnification never exceeds $\mu = 3$ across all possible source positions in the source plane \citep{Keeton_2005}. Therefore, adopting this maximum magnification factor of $\mu = 3$ leads to a minimum intrinsic SFR of $\sim 26.3 (41.7) \Msun \, \mathrm{yr}^{-1}$ for HD1 (HD2), which is still far too high, by more than a factor of 1.6 at the very least, compared to the theoretical prediction from the abundance of star-forming galaxies as a function of SFR by \cite{Pacucci_2022}. Therefore, simply adding plausible values of the ellipticity and external shear to the lensing mass model to obtain a single magnified image fails to account for the unusual luminosities of both sources. For extremely large values of ellipticity and shear ($\epsilon \geq 0.5$ and $\gamma \geq 0.3$), the maximum magnification for single images can in principle increase to infinity \citep{Keeton_2005}. However, galaxies with such extreme properties are incredibly rare. Previous studies on galaxy clusters produce ellipticity distributions with a mean of $\sim 0.3$ and a standard deviation of $\sim 0.2$ (\citealt{Jorgensen_1995}, \citealt{dEugenio_2015}). Meanwhile, $N$-body simulations to study the distribution of external shears arising due to neighbouring structures on a lensing galaxy, estimate that the mean external shear from  such neighbouring structures is $\gamma \sim 0.05$, with the probability of shears greater than $\gamma=0.1$ being a mere $\sim 16\%$ \citep{Holder&Shecter_2003}. While possible, such a lensing case requires a high degree of fine-tuning for both the properties of the lens and other masses in the vicinity. Therefore, we conclude that it is more likely that HD1 and HD2 have unusually high large luminosities that cannot be accounted for by this particular lensing scenario either.

\section{Strong Gravitational Lensing with Multiple Imaging and Spherical Deflectors}

We now turn our attention to the possibility that the $z \sim 13$ sources are strongly lensed and multiply imaged, without differentiating between the two distinct possibilities of the production of multiple images, namely: (i) the two magnified images are detected as a single resolvable image; (ii) only one of the multiple images is detected. These further detailed cases are addressed in Section \ref{sec:last}.

\subsection{Magnitudes of Hypothetical Lensing Galaxies}

We start by noting that in the imaging of HD1 and HD2 no evident sign of lensing is found, as the two sources appear to be isolated and point-like \citep{Harikane_2022_LBG}. To investigate the lensing hypothesis further, we now attempt to narrow our focus down to foreground lensing galaxies that are too dim to be observable, but still have a significant lensing effect on a $z \sim 13$ source. The limiting magnitudes of the imaging data used to identify HD1 and HD2 are provided in \cite{Harikane_2022_LBG}; the lowest limiting magnitude is $m=24.9$ in the 4.5 $\mu$m band images taken by the Spitzer Large Area Survey (SPLASH). With this in mind, we focus solely on the galaxies that have apparent magnitudes fainter than $m=24.9$ and study their potential lensing properties. In doing so, we are conservatively identifying as many galaxies that are too dim to be detected by the survey.

Here, we return our focus to lensing galaxies described by an SIS profile. For these putative lenses, we consider velocity dispersions in the range $1.6 \leq \log \sigma \leq 2.6$, in units of $\mathrm{km \ s}^{-1}$, and lens galaxy redshifts between $0.5 \leq z_L \leq 4$. We calculate the brightness of a galaxy at a given velocity dispersion and redshift using the Faber-Jackson relation \citep{Faber_Jackson_1976}. It is believed that the stellar mass-velocity dispersion relation evolves weakly with redshift beyond $z \sim 2$ \citep{Treu_2005}, and thus we use the evolving Faber-Jackson relation of the form $L \propto \sigma^4 (1+z_L)^\delta$ as determined by \cite{Mason_2015}. In their study, they determined the empirical form of the relation across redshifts, and obtain best-fit parameters for galaxies in the redshift ranges $z<0.5$, $0.5<z<1.0$, and $1.0<z<2.6$. Since we consider only galaxies between $0.5 \leq z \leq 4.0$, we use the empirical results for galaxies at $z>1.0$. Specifically, they obtain a fit of the form:
\begin{equation} \label{eq5}
    m = 10\left[ \log \left(\frac{(1+z)^a}{\sigma} \right) + b \right] \, ,
\end{equation}
where $m$ is the apparent AB magnitude of the galaxy, and the constants $a = 1.02 \pm 0.15$ while $b = 4.12 \pm 0.05$. We also note that the differences in $a$ and $b$ obtained between $0.5<z<1.0$ and $z>1.0$ are small, and thus our choice of $a$ and $b$ should adequately describe the relation between $m$ and $\sigma$ for galaxies across our whole range of consideration, namely, $0.5\leq z \leq 4$. In order to be as conservative as possible in our analysis, we take into consideration the uncertainties associated with $a$ and $b$ and use for our calculations the extreme values of $a = 1.17$ and $b = 4.17$. In doing so, we skew our calculations towards higher apparent magnitudes in order to consider as much as possible galaxies that are too dim to be detected.

\begin{figure*}
\includegraphics[width=1.4\columnwidth]{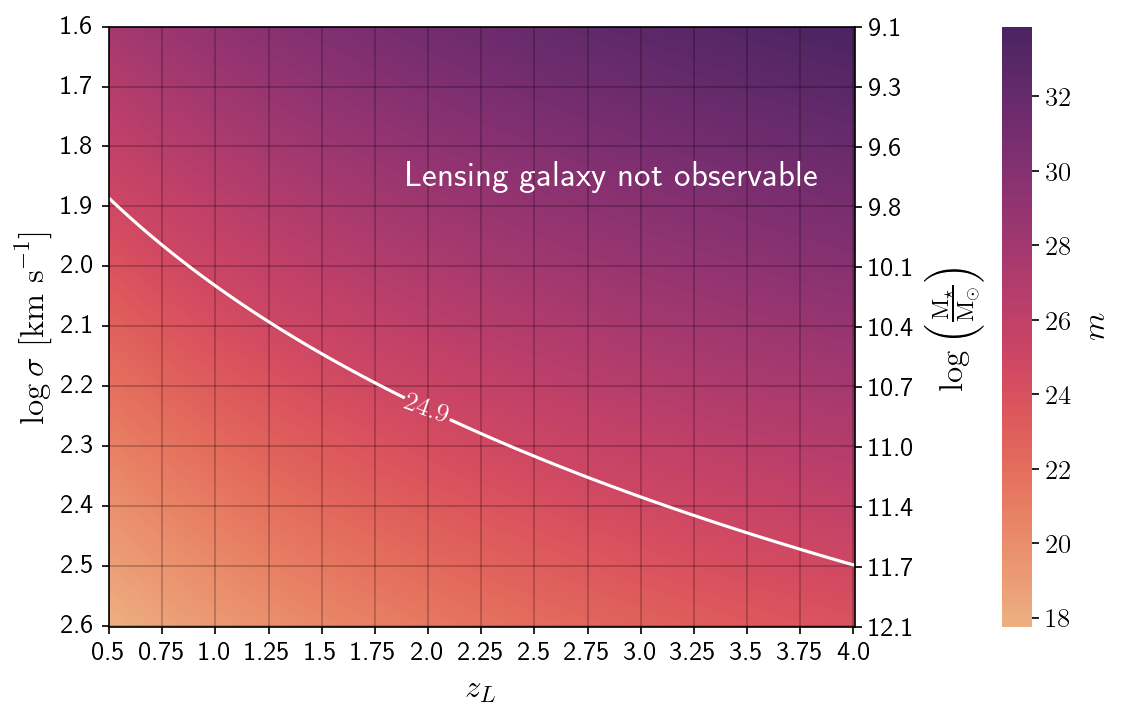}
\centering
\caption{Apparent magnitude of a hypothetical lensing galaxy, as a function of its velocity dispersion (or mass) and its redshift. The contour line indicates the level curve where the apparent magnitude is $m=24.9$. Plausible lensing galaxies bounded by the contour in the top right corner would be too faint to be detectable with current imaging.}
\label{fig:2}
\end{figure*}

The distribution of the apparent magnitude of a galaxy, $m$, across the whole parameter space in $\sigma$ and $z_L$ is shown in Fig. \ref{fig:2}. Using the $M_\star-\sigma$ relation from \cite{Zahid_2016}, we display the stellar mass of each galaxy alongside the velocity dispersion. The white contour separates the magnitude distribution into two regions, with the region above the contour representing fainter galaxies that have apparent magnitude more than 24.9, and thus are undetectable.

Implicit from the fact that HD1 and HD2 appear isolated is the condition that we only consider the parameter space of isolated field galaxies. In the context of observed galaxy evolution, the stellar mass function at relatively higher redshifts of $z>2$ drops drastically past $M_{\star} \sim 10^{11} \Msun$ (see, e.g., \citealt{Fontana2006}, \citealt{Marchesini_2009}). We then expect the number of unobservable galaxies within the parameter space $z>2$ and $M_{\star} > 10^{11} \Msun$ to be significantly lower, suggesting that the real lensing probability might be lower than what subsequent calculations might suggest.

\subsection{Multiple Imaging Optical Depth}
\label{subsec:tau}

Having constrained the parameter space of potential lens galaxy properties, we now investigate the possibility that the $z \sim 13$ sources are strongly lensed and magnified to produce multiple images, by computing the lensing optical depth. Given a source at redshift $z_S$, the optical depth, $\tau$, is the fraction of the sky covered by the cross section within the caustics of all foreground galaxies at redshift $z_L$, summed over all redshifts up to $z_S$ along all lines of sight. 
Assuming little to no overlap of the different caustics, $\tau$ is the cross-section for the source galaxy to lie within the caustics of a particular lens galaxy, and can thus be interpreted as the probability that the source galaxy at redshift $z_S$ is strongly lensed, producing multiple images. 

In order to keep this computation analytic, we further assume that all foreground lens galaxies are described by SIS profiles, since considering the lensing cross sections of asymmetric mass distributions greatly complicates calculations \citep{Blandford_1997, Keeton_1997}. For our purposes, disregarding possible variations in ellipticity and shear across all foreground deflectors is reasonable, since they only serve to decrease the multiple imaging cross section calculated from SIS deflectors, and in turn the optical depth by as little as $\sim 0.6 \%$ \citep{Huterer_2018}. Thus, we focus only on varying $\sigma$ and $z_L$ such that the parameter space we have to study is limited.

By considering only SIS lenses, the multiple image cross section is the area within the Einstein radius of the lensing configuration. \cite{Wyithe_2011} and \cite{Mason_2015} calculate the lensing optical depth as:
\begin{equation} \label{eq6}
    \tau(z_S) = \int_0^{z_S} dz_L \int d\sigma \Phi(\sigma, z_L) (1+z)^3 c \frac{dt}{dz_L} D_L^2 \pi \theta_E^2(\sigma,z_L) \, ,
\end{equation}
where $\Phi(\sigma, z_L)$ is the velocity dispersion function of the foreground lens galaxies, and $dt/dz_L = 1/[(1+z_L)H(z_L)]$, where $H(z_L)$ is the Hubble parameter at $z_L$ \citep{Carlstrom_2002}. Since we are assuming SIS mass profiles for the deflectors, $\theta_E$ is the usual Einstein radius for an SIS profile given by Eq. \ref{eq2}. We use the redshift dependent velocity dispersion function obtained by \cite{Mason_2015}, which describes the number density of deflectors as a function of velocity dispersion $\sigma$ and deflector redshift $z_L$:
\begin{equation} \label{eq7}
    \Phi(\sigma,z_L) = p^{-1} \frac{\Phi^*(z_L)}{\sigma (1+z_L)^{\delta}} \left( \frac{\sigma}{\sigma^*} \right)^{p^{-1} (1+\alpha)} \exp{\left[ - \left( \frac{\sigma}{\sigma^*} \right)^{p-1} \right]} \, ,
\end{equation}
where $p = 0.24 \pm 0.02$, $\delta = 0.20 \pm 0.07$, $\Phi^*(z_L) = (3.75 \pm 2.99) \times 10^{-3} (1+z_L)^{-2.46 \pm 0.53} \ \mathrm{Mpc}^{-3}, \alpha = -0.54 \pm 0.32$ and $\sigma^* = 216 \pm 18  \ \mathrm{km \ s}^{-1}$. 

For subsequent calculations involving Eq. \ref{eq7}, we find that the behaviour of $\Phi(\sigma,z_L)$ becomes unpredictable as we take into consideration the variation of each parameter across their uncertainty ranges given above. This is particularly evident for the uncertainty in $p$, whose variation across its extremal values is not monotone, but instead results in an uneven redistribution of $\Phi$ across the entire parameter space. For a rough sense of this uneven variation, we find that increasing $p$ from $p=0.22$ to $p=0.26$ causes $\Phi$ to increase for galaxies roughly in the range $250 \ 
\mathrm{km \ s}^{-1} \leq \sigma \leq 400 \ \mathrm{km \ s}^{-1}$ (these high velocity dispersions are measured for extremely massive galaxies including central galaxies in clusters) and decrease for galaxies elsewhere. We therefore choose to disregard the empirical uncertainty associated with each parameter in Eq. \ref{eq7} given above, and use the central values for all subsequent calculations. By considering the quantity $P(\sigma,z_L)$, defined formally in the upcoming Eq. \ref{eq8}, as the relative contribution towards $\tau$ by a particular galaxy with $\sigma$ and $z_L$, we note that the deviation of $\Phi$, and consequently $P$, from their central values is significant when the parameters are varied across their uncertainty ranges. For reasons that will become clear later on, we find that the variation in $P$ under specific circumstances is negligible, further justifying our choice to disregard the uncertainties in the parameters.

With the above assumption, we now calculate the total optical depth for HD1 at $z=13.27$ as $\tau \approx 0.00161$. We then consider the contribution to the total optical depth $\mathrm{d}^2 \tau/\mathrm{d}\sigma \mathrm{d} z_L$, given by the integrand in Eq. \ref{eq6} as a function of lens redshift and velocity dispersion to explore the parameter space for lensing galaxies. Specifically, we calculate the ratio of each individual contribution over the total optical depth:
\begin{equation} \label{eq8}
    P(\sigma, z_L) = \left. \frac{\mathrm{d}^2 \tau}{\mathrm{d}\sigma \mathrm{d} z_L} \middle/ \tau \right.
\end{equation}
It follows from Eq. \ref{eq6} and \ref{eq8} that, for a given $\tau$, 
\begin{equation}
    \int_0^{z_S} \int P \ dz_L d\sigma = 1 \, .
\end{equation} 
Thus, if the source is lensed by a foreground galaxy, $P\, dz_L d\sigma$ approximates the probability of a galaxy at a particular redshift and velocity dispersion being the foreground lens. With this, $P(\sigma, z_L)$ serves as a good measure of the relative likelihood of a galaxy being the lens, independent of the value of $dz_L$ and $d\sigma$, provided that they are small enough.

\begin{figure*}
\includegraphics[width=1.4\columnwidth]{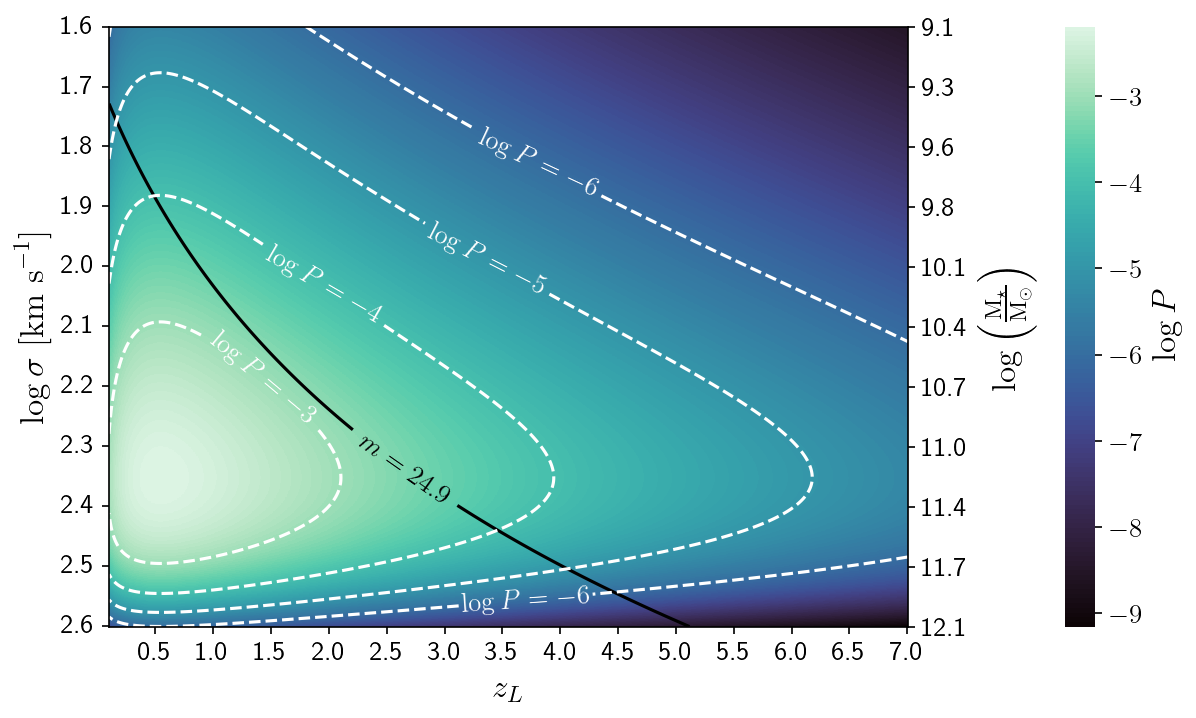}
\centering
\caption{Distribution of $\log P$ which measures the relative contribution of a particular galaxy to the lensing optical depth, as a function of its velocity dispersion (or mass) and its redshift. The white contours indicate the level curves for $\log P = -3, -4, -5$ and $-6$. The galaxies bounded by the black contour to the right have apparent magnitudes $m > 24.9 $, and are thus undetectable with current imaging \citep{Harikane_2022_LBG}.}
\label{fig:3}
\end{figure*}

The distribution of $\log P$ as a function of $\log \sigma$ and $z_L$ is shown in Fig. \ref{fig:3}, with the same $m=24.9$ contour from Fig. \ref{fig:2}. Here, the range of redshifts depicted in the figure is enlarged as compared to Fig. \ref{fig:2} to $0.5 \leq z_L \leq 7.0$ in order to show the full extent of levels in the curves of $\log P$.

Given a hypothetical lensing galaxy and integrating $P$ over the area bound within the relevant contour lines, we expect to find the lens in the parameter space bounded within the $\log P = -4$ with a $~97.23 \%$ probability (corresponding to a $\sim 2 \sigma$ confidence level). Similarly, we expect to find the lensing galaxy in the parameter space within $\log P = -5$ with a $99.66 \%$ probability ($\sim 3 \sigma$ confidence level).

These calculations indicate that a lensing galaxy, if any, can be found within the $\log P = -5$ $(\log P = -4)$ parameter space at a $3\sigma$ ($2\sigma$) confidence level. However, the galaxies that are the most likely lensing candidates for HD1 are precisely the ones that are within the region bounded by the $m=24.9$, since they are the galaxies that are unobservable with current surveys. With this, we assert that galaxies in the parameter space bounded by the $m=24.9$ curve and the $\log P = -5$ ($\log P = -4$) curve are the likely candidates for foreground galaxies lensing the light from HD1 at a $3\sigma$ ($2\sigma$) confidence level. Furthermore, integrating the distribution in $P$ across the parameter space for which $m \geq 24.9$, we find that in the event that HD1 is indeed lensed by some hypothetical galaxy, the probability of the galaxy having an apparent magnitude $m \geq 24.9$ and thus being too faint to be observable by the current survey, is $7.39 \%$. Equivalently, this asserts at a $92.61 \%$ confidence level that there is no SIS galaxy across the whole parameter space that is lensing HD1. Performing the calculation for deep Hubble Space Telescope (HST) imaging surveys with a limiting magnitude of $m_{\rm lim} \sim 30$ \citep{Illingworth_2013}, we find that there is a mere $0.058 \%$ theoretical probability of the hypothetical lensing galaxy being too faint to be detected, while for the JWST with $m_{\rm lim} \sim 34$, the theoretical probability would drop down even further to $0.0012 \%$.

To conclude, we note that the shapes of the white contours drawn in Fig. \ref{fig:3} are determined primarily by the behaviour of $\Phi$, the number density of deflectors at a particular redshift and velocity dispersion. Since the parameter $P$ represents the probability of a galaxy at a particular redshift and velocity dispersion being a possible lensing candidate, the distribution of $P$ depicted in Fig. \ref{fig:3} indicates how effective of a lens any particular galaxy is, as well as the likelihood that the galaxy is indeed lensing HD1. We expect this to be relevant in informing the observational search for possible lensing galaxies.

Furthermore, by focusing on the region of the parameter space between the $m=24.9$ and the $\log P = -5$ curves, we find that the variation in $\log P$ by varying the parameter $p$ in Eq. \ref{eq7} across its uncertainty range is at most $6 \%$. This justifies our earlier decision to disregard the uncertainty of parameters in Eq. \ref{eq7}.

\section{Strong Lensing with Elliptical Deflectors}
\label{sec:last}
We conclude by studying the possible lensing scenarios that can account for the unusually high UV luminosity observed for HD1 while differentiating between multiple images that are distinct, and multiple images that are resolveable only as a single image. Furthermore, we are now interested in lensing scenarios that produce images with sufficiently high total magnification. In practice, we obtain the total magnification $\mu_{\mathrm{tot}} = \Sigma_{i} \vert \mu_i\vert$ of the multiple images produced by a lensing scenario and use Eq. \ref{eq1} to infer the intrinsic SFR of HD1. We are interested in looking for lensing scenarios corresponding to inferred SFRs that are less than or equal to the theoretical expectations of $16.5 \Msun \, \mathrm{yr}^{-1}$, extrapolated from lower redshifts for both HD1 and HD2 \citep{Pacucci_2022}.

Considering that HD1 and HD2 are observed to be point-like, the multiple images produced by any lensing scenario need to be sufficiently close to be perceived as single, due to the finite angular resolution of the telescopes used. Since HD1 and HD2 were detected with ground based telescopes this is an interesting and important case to explore. Across the different ground-based surveys used to detect HD1 and HD2, as listed in \cite{Harikane_2022_LBG}, the median resolution across all bands in each survey is $\sim 0.6$ arcsec FWHM  (\citealt{Aihara_2018}; \citealt{McCracken_2012}; \citealt{Lawrence_2007}). We therefore filter out plausible lensing scenarios by requiring that the lens produce all pair-wise image separations to be $<0.6$ arcsec. Since the quoted resolution is merely a median value across all bands, it may very well be an underestimation of the true FWHM in the images of HD1 and HD2. We expect that an increase in the FWHM allows for more images that have a larger pairwise angular separation to be detected as a single image by current observations. This will loosen the criteria for possible lensing scenarios that still produce a single image matching current observations of HD1 and HD2 as point-like sources, thus increasing the overall probability of lensing in subsequent calculations.

In contrast to the previous sections, now we also consider deflector galaxies with spherically asymmetric lensing potentials to consider cases when single images with magnifications much higher than the SIS limit of $\mu = 2$ are produced. Our study would then be more realistic, consistent with the fact that most lensing galaxies have non-spherical mass distributions, and that $\sim 80 \%$ of lenses are more appropriately described by singular isothermal elliptical mass (SIE) models (\citealt{Kochanek_1996}; \citealt{Keeton_1997}). In order to limit the size of the parameter space exploration for computational feasibility, we focus only on asymmetries resulting from the elliptical mass distribution of deflector galaxies and ignore the contribution of external shear, as both HD1 and HD2 appear to be isolated from nearby objects. 

With the above requirements in mind, we study various lensing scenarios across a range of SIE deflector galaxy properties parametrized by $\sigma,z_L$ and $\epsilon$. Specifically, we use the \href{https://pyautolens.readthedocs.io/}{\color{cyan} PyAutolens} package to simulate gravitational lensing systems with a point source located in the source plane at $z_S = 13.27$ \citep{pyautolens, Nightingale2015, Nightingale2018}. The deflector galaxy with velocity dispersion $\sigma$ and ellipticity $\epsilon$ is placed on the image plane at redshift $z_L$. For each set of parameters specifying a given deflector galaxy, we solve the lens equation for the image of the point source while varying the source position across an appropriately-sized source plane grid to obtain sets of image positions and their associated magnifications. 

The set of all possible source positions is then culled based on the following criteria in order of execution: (i) The separation between any pair of images that are not sufficiently demagnified must be less than $0.6$ arcsec. Using the apparent magnitude of HD1, $m = 24.6$ in the 3.6 $\mu$m band, we find that in the extreme case where HD1 is not lensed and the apparent magnitude measured above is intrinsic, isolated images that have magnifications $\mu < 0.525$ will have apparent magnitudes that are demagnified below the limiting magnitude of both the 3.6 and 4.5 $\mu$m bands. Therefore, if the separation of any pair of images is more than $0.6$ arcsec but either image has $\mu < 0.525$, the faint isolated image will be undetectable by the survey. Thus, our physical criteria that only a single distinct image is observed by the survey is not violated, and the source plane points which the particular image distribution belongs to is not rejected outright. (ii) The SFR of the source as determined by obtaining the net magnification of all images and using Eq. \ref{eq1} must be less than the upper bound of $16.5 \Msun \, \mathrm{yr}^{-1}$. Afterwards, we count the number of source plane points that satisfy our filter criteria to obtain the cross section, which we denote by $A$, of the source plane that corresponds to lensing scenarios which account for the unusually high luminosity of HD1. This procedure above was repeated for different lensing galaxies across the parameter space spanned by the respective ranges in velocity dispersion and ellipticity: $1.6 \leq \log \sigma \leq 2.6$, $0.1 \leq z_L \leq 7$ and $0 \leq \epsilon \leq 0.9$. Since the strength of lensing only depends on $\sigma$ and $z_L$ indirectly through the Einstein radius $\theta_E(\sigma,z_L)$, we can effectively reduce the parameter space down to two dimensions along $\theta_E$ and $\epsilon$, which now extends over $0.022 \leq \theta_E \mathrm{[arcsec]} \leq 4.377$ and $0 \leq \epsilon \leq 0.9$ respectively. Thus, we sample the variation of $A$ over the parameter space across 160 points (see additional comments in the Appendix). 

We assume that the lensing cross sections $A$ corresponding to different galaxies do not overlap, and calculate the lensing optical depth due to SIE galaxies in such a way that accounts for its unusually high observed luminosity as (see, e.g., \citealt{Keeton_2005, Huterer_2018}, and our Eq. \ref{eq6}):
\begin{equation} \label{eq10}
\begin{split}
    \tau_A= & \int_0^{z_S} dz_L \int d\sigma \Phi(\sigma, z_L) (1+z)^3 c \frac{dt}{dz_L} D_L^2 \times \\ 
    & \int_0^{1} d\epsilon p(\epsilon) A(\sigma, z_L, \epsilon) \, .
\end{split}
\end{equation}
Differently from Eq. \ref{eq6}, we now integrate the quantity $A(\sigma, z_L, \epsilon)$, which is the cross section of the source plane leading to image configurations that satisfy our filtering criteria above. We note that the dependence of Eq. \ref{eq10} on the Einstein radius manifests itself through $A$, which indirectly depends on $\theta_E$ through $\sigma$ and $z_L$. In addition, we also integrate over a realistic distribution of ellipticities in an arbitrary population of galaxies, $p(\epsilon)$, where we have used the empirically measured ellipticity distribution in \cite{dEugenio_2015}. 

With the data obtained from the simulations, we calculate the modified optical depth in Eq. \ref{eq10} as $\tau_A \approx 5.38 \times 10^{-5}$. Formally, this is not the total probability that the unusually high luminosity of HD1 can be accounted for through gravitational lensing, because this value is calculated by integrating over a fraction of the parameter space.

However, we assert that the calculated value of $\tau_A$ is sufficiently close to the true probability for the following reasons: (i) We expect the integral over $0.9 \leq \epsilon \leq 1$ to contribute negligibly to the probability, owing to the extremely low value of $p(\epsilon)$ for $\epsilon \geq 0.9$. (ii) Considering the $z_L$ and $\sigma$ dependence of $\tau_A$ indirectly by focusing on $\theta_E$, we can bound the contribution towards the total probability from the "missing" parameter space in $\sigma$ and $z_L$ that was not accounted for in our simulations. Over portions of the parameter space where $\theta_E$ is less than the maximum simulated value of $\theta_{\mathrm{max}} \approx 4.377 \ \mathrm{arcsec}$, we calculate $\tau_A$ using the measured value of $A$ corresponding to $\theta_{\text{max}}$. Since $A$ scales with $\theta_E$, we expect the true values of $A$ over the missing parameter space to be much lower than the value of $A$ corresponding to $\theta_{\text{max}}$. Thus, $\tau_A$ calculated here is generously overestimated. For the other portion of the parameter space where $\theta_E > \theta_{\mathrm{max}}$, we use unrealistically high values of $A$ much larger than the expected values (around the size of the multiple imaging cross section for an SIS lens at any given $\theta_E$) to calculate $\tau_A$. We then find that the maximum contribution from the missing parameter space is at most $4.82 \times 10^{-7}$, which is $\sim 0.9 \%$ of our fiducial value of $\tau_A \approx 5.38 \times 10^{-5}$. In fact, we expect this contribution to be much lower, owing to the significant over-estimations of $A$ assumed in our calculations. We also expect a slight overestimation of our fiducial value due to grid sampling limitations in the simulation (see the Appendix). Despite this, the argument above lets us place an upper bound of $\sim 5 \times 10^{-5}$ on the total modified optical depth, or equivalently, $\sim 0.005 \%$ on the total probability that HD1 is lensed in such a way that accounts for its unusually high luminosity. Equivalently, our results assert at $ > 99.99 \%$ confidence level that HD1 is not lensed by a SIE deflector in a way that accounts for its unusually high luminosity.

Although the lensing hypothesis is disfavored, in the unlikely event that HD1 is lensed by an SIE deflector to produce a single resolvable image that is sufficiently magnified, we can still study the regions of the parameter space where such hypothetical lens galaxies are most likely to exist. Using the upper bound for $\tau_A$ calculated above, we can repeat a similar analysis as in the case of the multiple imaging optical depth in \S \ref{subsec:tau} and calculate the contribution to the total lensing probability $P_A$, given by the ratio of the integrand in Eq. \ref{eq10} over $\tau_A$:
\begin{equation} \label{eq11}
    P_A(\sigma,z_L) =\left. \frac{\mathrm{d}^2 \tau_A}{\mathrm{d}\sigma \mathrm{d} z_L} \middle/ \tau_A \right. \, .
\end{equation}

\begin{figure*}
\includegraphics[width=1.4\columnwidth]{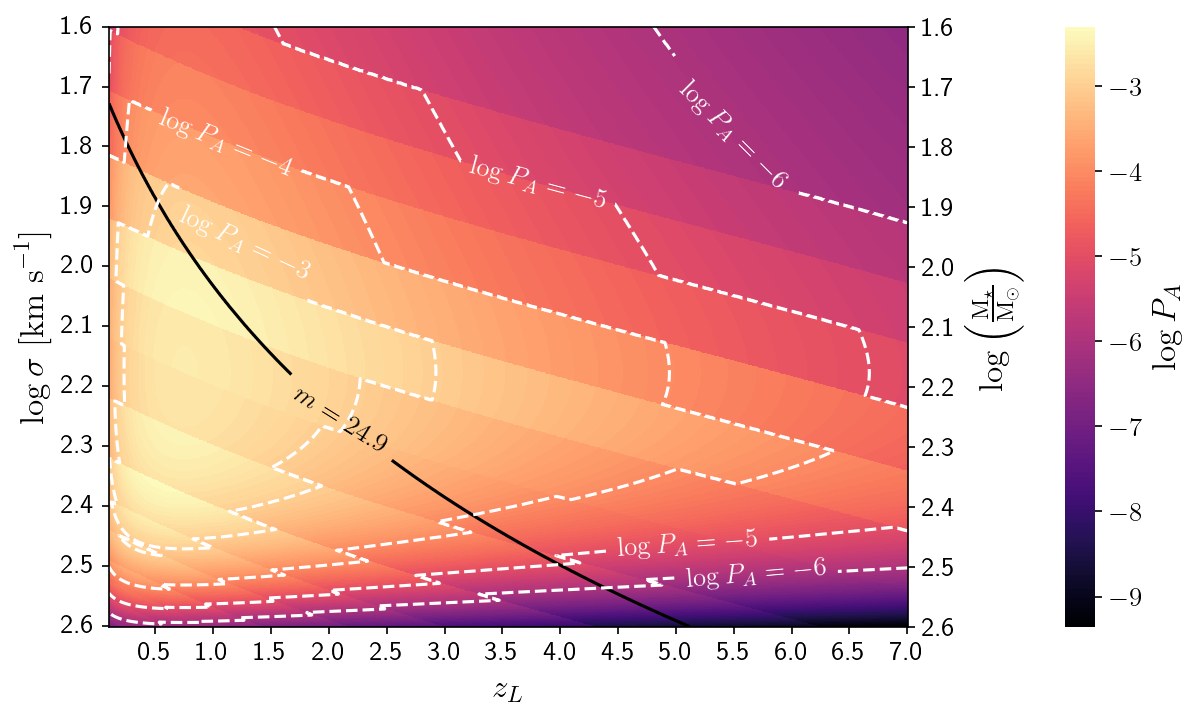}
\centering
\caption{Distribution of $\log P_A$ which measures the relative contribution of a particular galaxy to the modified lensing optical depth $(\tau_A)$, as a function of its velocity dispersion (or mass) and its redshift. The white contours indicate the level curves for $\log P_A = -3, -4, -5$ and $-6$. The galaxies bounded by the black contour to the right have apparent magnitudes $m > 24.9 $, and are thus undetectable with current imaging \citep{Harikane_2022_LBG}.}
\label{fig:4}
\end{figure*}

$P_A(\sigma,z_L)$ serves as a good measure of the relative likelihood that a galaxy at a redshift $z_L$ with velocity dispersion $\sigma$ is the one responsible for lensing. We note that implicit in our definition of $P_A$ in Eq. \ref{eq11} is an integral of all possible ellipticities of any particular galaxy over the ellipticity distribution mentioned previously. In other words, $P_A$ readily accounts for the weighted contribution from all ellipticities towards the probability that a particular galaxy is the deflector responsible for lensing. With this, the distribution in $P_A$ against $\log \sigma$ and $z_L$ is shown in Fig. \ref{fig:4}.
Here, we choose to limit our calculation of $P_A$ to the parameter space spanned by $1.6 \leq \log \sigma \leq 2.6$ and $0.1 \leq z_L \leq 7$, since we do not have an accurate measure of the distribution in $P_A$ outside of this range.

We then expect to find the hypothetical lensing galaxy in the area within $\log P = -4$ with a probability of $96.1 \%$ (corresponding to slightly more than a $2\sigma$ confidence level), while the probability turns out to be $98.9 \%$ for $\log P = -5$ ($\sim 2.5\sigma$ confidence level). Further taking into account the limiting magnitude of the current survey at $m_{\mathrm{lim}} = 24.9$, we find that in the event that HD1 is indeed lensed by a SIE deflector in a manner that accounts for its unusually high luminosity, the probability of the deflector galaxy being too faint to be observable by the current survey is $30.9 \%$. For the JWST (HST) limiting magnitude of $m_{\lim} \sim 34 \, (30)$, the probability of the hypothetical lensing galaxy being unobservable drops significantly to $0.0025\%$ ($0.245 \%$). 

\section{Discussion and Conclusions}

This work was motivated by the discovery of two candidate galaxies at $z\sim13$, HD1 and HD2 \citep{Harikane_2022_LBG, Pacucci_2022}. Considering that they appear extremely UV bright, we address the hypothesis that their observed luminosities can be explained through gravitational lensing. In particular, we consider three different lensing cases, with constraints from current observations of HD1, namely: (i) Sufficiently magnified single imaging by SIS deflectors; (ii) Multiple imaging by SIS deflectors; and (iii) Sufficiently magnified, single resolvable imaging by SIE deflectors. Our main results can be summarized as follows:
\begin{itemize}
  \item In the single-image regime assuming only SIS deflectors, the single images produced by the lens can never be sufficiently magnified to account for the unusually high observed luminosity of HD1. For asymmetric lensing potentials, the maximum magnification for single images only exceeds a value of 3 for relatively high values of the ellipticity of the mass distribution $\epsilon > 0.3$ and the external shear $\gamma >0 .1$. For typical ellipticities and external shear amplitudes, the magnification cannot attain sufficiently high values to account for the unusual luminosity of HD1.
  \item In the multiple-image regime assuming only SIS deflectors, the lensing optical depth, which approximates the probability that HD1 is lensed to produce multiple images, is $\tau \approx 0.002$. If HD1 is lensed by an SIS deflector to produce multiple images, the probability that the deflector is too faint to be observable by current imaging is $7.39 \%$. If the HD1 field were to be surveyed instead using the JWST (HST) imaging, the probabilities would drop to $0.0012 \%$ ($0.058 \%)$. This suggests that with current observational data, there is a high likelihood of $>90 \%$ that HD1 is not lensed by an SIS deflector.
  \item Considering both the singly magnified and the magnified and multiply imaged lensing regimes for SIE deflectors, we calculate the probability that HD1 is lensed to produce images that are resolvable only as a single image by the current ground based surveys that detected HD1 and HD2, while still magnifying the observed image sufficiently to account for the unusual luminosity of HD1. Our calculations show that the modified lensing optical depth, representing the probability that the unusual luminosity of HD1 can be explained by gravitational lensing, has an upper bound of $5.42 \times 10^{-5}$. This corresponds to a $\sim 0.005 \%$ probability, suggesting that it is highly unlikely that the lensing hypothesis can explain the unusual properties of HD1. In the implausible scenario where the lensing hypothesis might hold true, there is a relatively high chance at $30.9 \%$ that the SIE deflector is too faint to be observable with the current survey. Using JWST (HST) imaging limiting magnitudes, the probability drops to $0.0025 \%$ ($0.245 \%$).
\end{itemize} 

In our work, we have assumed a single lens plane in all lensing scenarios. With high-redshift sources, of which HD1 and HD2 are prime examples, the probability for there to be multiple galaxies at different redshifts involved in the lensing scenario increases. In the case that the different lenses involved are aligned, but their redshifts are different, the simplest models for multiply-lensed systems allow for the individual magnifications produced by each lens to be summed up linearly. Here, we have not considered the possibility that multiple galaxies, each of which would magnify HD1 and HD2 insufficiently, could be aligned in such a way to give rise to a total magnification that would be high enough to account for their unusual luminosity. In practice, however, we expect that requiring two or more galaxies to line up along the line of sight to give rise to the effect above dramatically reduces the probability of such lensing scenarios. Indeed, at source redshifts up to $z_S =10$, the lensing optical depth for double lensing is less than that of single lensing by about a factor of 10 \citep{Moller_2001}.

While our findings point to the lensing hypothesis being extremely unlikely, in the exceptionally improbable scenario where HD1 is indeed lensed by a hypothetical foreground galaxy, our results do not rule out at a significant confidence level the possibility of there being such galaxies that are too faint to be detected by current imaging. At $7.39 \%$ and $30.9 \%$ respectively, the probability that a faint SIS/SIE deflector continues to remain undetected is too statistically significant to allow us to completely preclude the possibility that the unusual properties of HD1 and HD2 are simply a result of an undetected gravitational lensing event. What our results also show, however, is that such hypothetical faint objects cannot continue remaining hidden if we observe these galaxies with more sensitive instruments. In its upcoming observations of HD1 and HD2, JWST, with its limiting magnitude of $m_{\lim} \sim 34$, will be able to rule out at a more than $99.99 \%$ confidence level the possibility of there being a lensing galaxy that is still too faint to be observable. Combined with additional spectroscopic observations that will test the tentative redshifts of both sources, JWST will then be able to shed light on the physical nature of HD1 and HD2, opening new possibilities for studying galaxy and black hole formation in the earliest periods of the Universe.

\section*{Acknowledgements}

R.Z.L. acknowledges support from the Harvard College Research Program (HCRP).
F.P. acknowledges support from a Clay Fellowship administered by the Smithsonian Astrophysical Observatory. F.P., P.N. and A.L. gratefully acknowledge support from the Black Hole Initiative at Harvard University, which is funded by grants from the John Templeton Foundation and the Gordon and Betty Moore Foundation.

\section*{Data Availability}

The code used to analyze the lensing configurations and their likelihoods will be shared on reasonable request to the corresponding authors.



\bibliographystyle{mnras}
\bibliography{ms} 




\section*{Appendix: Details on the calculation of the lensing optical depth for SIE case}

In the simulations to compute the lensing optical depth in the elliptical case, $A$, there are a few interesting points that we note below. First, in contrast to \cite{Huterer_2018} and \cite{Keeton_2005}, we find that the dimensionless cross section, $A/\theta_E^2$, is not independent of $\theta_E$, and by association $\sigma$ and $z_L$. Our finding differs from these prior calculations due to the additional constraint on the separation of the images that we impose. Computing $A$ as described in Sec. \ref{sec:last}, by filtering out images that are too far apart, we irreversibly alter the dependence of $A$ on the physical scale of the lensing system, characterized by $\theta_E$ as we introduce a second scale in the problem — the image separation. As $\theta_E$ increases, the typical deflection angle of the lensing system scales in tandem and the same proportions of source positions will on average split into image positions that have a larger angular separation. Therefore, it is not surprising that in the case considered,  where we are only interested in images separated below a fixed threshold, even if $A$ increases with $\theta_E$ simply due to larger lensing cross sections, $A/\theta_E^2$ instead decreases.

Second, the degree of precision with which we vary the position of the point sources — our sampling of the source plane — in the simulations is severely limited by computation time. At larger $\theta_E$, the point sources differ in position by $0.01$ arcsec between each iteration. Using a smaller steps, say $0.001$ arcsec, increases the run time of the simulation by a factor of $\gtrsim 10$. As we decrease $\theta_E$ towards the other end of the parameter space, however, we find that using the same precision ($0.01$ arcsec per point source) severely impacts the accuracy of the calculated value of $A$ as well. For $\theta_E < \sim 0.05$ arcsec, the fraction of the multiply imaged cross section (i.e., the area inside the caustics of the lensing system) occupied by a single point in the source plane is $\gtrsim 10 \%$. In this case, we expect that near the edges of the cross section in the source plane that fit our criteria for $A$, the relatively large point sources which we use as markers to indirectly measure this cross section spill over to a large extent into the region of the source plane that does not fit our criteria in measuring $A$. Thus, we expect there to be a significant over-estimation for the value of $A$ at smaller $\theta_E$. 

Furthermore, we observe that due to limitations in the numerical precision in solving of the lens equation in \href{https://pyautolens.readthedocs.io/}{\color{cyan} PyAutolens}, there were occasions where multiple images inconsistent with the lensing theory (e.g., images with high magnifications that were far away from the critical lines) were returned by the code. Even though attempts were made to identify and remove most of these erroneous points where possible, we were unable to implement a consistent way for our script to automatically detect and remove all of the possibly erroneous images. Therefore, we expect a further overestimation of $A$ originating from the inclusion of some source points that passed the filter criteria due to the presence of the erroneous points. While it is difficult to evaluate how much $A$ is overestimated as a result of the above factors, it is still worth noting that all quantities calculated using $A$ will be an upper bound that includes the overestimation in $A$.


\bsp	
\label{lastpage}
\end{document}